# Automation of gene function prediction through modeling human curators' decisions in GO phylogenetic annotation project

Haiming Tang[1,*], Paul Thomas[1,2], Huaiyu Mi[1]

1. Department of Preventive Medicine, University of Southern California, Los Angeles, CA

2. Department of Biological Sciences, University of Southern California, Los Angeles, CA

* Corresponding author

## Abstract

The Gene Ontology Consortium launched the GO-PAINT project (Phylogenetic Annotation and INference Tool) 9 years ago and is currently being used in the GO Reference Genome Annotation Project to support inference of GO function terms (molecular function, cellular component and biological process) by homology. PAINT uses a phylogenetic model to infer gene function by homology, a process that requires manual curation of experienced biocurators. Tremendous amount of time and efforts have been spent on the GO-PAINT project yielding more than 4000 fully annotated phylogenetic families with more than 170,000 annotations. These preliminary data have thus enabled potential algorithmic representation and automatic solvation of the additional 9000-unannoated phylogenetic families. Here we present an automated pipeline for phylogenetic annotation and inference, which simulates the standard annotation procedures of curators and models the curators' decisions during the manual curation process. The pipeline has been built into the newest version of PAINT software available at http://www.pantherdb.org/downloads/index.jsp. The standalone automation pipeline and datasets are available at https://github.com/haimingt/GO-PAINT-automation

# Introduction

Pathway analysis of genomic data—the use of prior knowledge about how genes function together in biological systems—plays an increasingly critical role in gaining biological insights from large-scale genomic studies, and particularly in cancer research. One of the more recent applications of pathway analysis has been in GWAS studies, where it has been used not only to suggest pathways involved in disease etiology, but also to discover other potential genetic factors acting in common pathways that did not reach genome-wide significance when considered in isolation. Some high-profile examples include breast cancer (Fachal and Dunning 2015; Kraft and Haiman 2010), Crohn's disease (Mathew 2008) and type 2 diabetes (Sanghera and Blackett 2012). Pathway analysis leverages prior knowledge of which genes act together, and thus may have similar effects when mutated or otherwise perturbed. The richest source of such information, in a computer-accessible form is the annotation of genes with biological process terms (which include pathways) from the Gene Ontology (GO) (Ashburner, et al. 2000; The Gene Ontology 2017). However, even GO annotations are very incomplete. This incompleteness dramatically impairs our ability to use prior pathway knowledge in genomic analyses of human data, including identifying cancer risk, mechanism, and treatment.

The most promising approach for increasing the completeness of GO annotations is to infer gene function from the extensive studies on related genes in the highly studied "model" organisms: the mouse, zebrafish, fruit fly, nematode worm, budding yeast, fission yeast, the slime mold Dictyostelium, the plant Arabidopsis and the bacterium E. coli. The challenge lies in determining exactly which functions are evolutionarily conserved, and among which groups of organisms. Recent work in the GO Consortium, the GO Consortium Phylogenetic Annotation Inference Tool Project (GO-PAINT) provides an estimate of how much more functional information can be inferred for human genes using such an approach(Gaudet, et al. 2011). In the gene families annotated so far, inclusion of model organism-based inferences increased the number of unique human gene annotations by about 5-fold overall (from 2068 annotations to 10718, for 2243 genes), and the number of biological process annotations by almost 6-fold (from 900 to 5288). It is probably useful to illustrate the problem with a concrete example: the one-carbon pathway via folate as a carrier (GO:0006730, one-carbon metabolic process). Defects in this pathway have been implicated in many disorders including neural tube defects (Beaudin and Stover 2007; Ducker and Rabinowitz 2017) and colon cancer risk (Duthie 2011; Locasale 2013; Yang and Vousden 2016). Currently there is only one human gene annotated to this pathway in the GO database. But there are numerous genes annotated from experimental evidence in other organisms: 12 in budding yeast, 4 in fruit fly, 4 in the fungus Aspergillus nidulans, 1 each in mouse, rat and fission yeast. These model organism genes all have homologs in the human genome, and thus could be used to infer the functions of the related human genes.

To date, two "genome-scale" methods have been developed to address the general GO term inference problem. The first, implemented in Blast2GO (Conesa and Gotz 2008; Conesa, et al. 2005) uses NCBI BLAST (Johnson, et al. 2008) to compare each un-annotated input sequence to a database of sequences of proteins that have been annotated with GO terms. The input sequence is then annotated by transferring the GO terms of each annotated sequence that "hits" the input sequence above some threshold. This threshold can include more than one metric from the BLAST alignment such as statistical significance (E-value) and alignment length. Nevertheless, and this is the crucial point, the BLAST threshold is the same for all proteins and all functions. This is obviously an extremely crude approximation: the sequence similarity of protein pairs with the same functions is extremely variable depending on the protein family, and, even more crucially, on the particular function in that family. In other words, the function inference problem can only be solved accurately on a case-by-case basis, for which any pairwise method is poorly suited. More recently, a more sophisticated inference method has been implemented by the Ensembl Compara project. The method itself is not published, but is described on the web (http://www.ebi.ac.uk/GOA/compara_go_annotations) and is currently supplying many inferred GO annotations. This is a phylogenetic method, in that it uses gene trees, so it avoids the problems arising from the uniform threshold in Blast2GO. Although it does not attempt to reconstruct ancestral states, its inferences could in principle be expressed as a particularly conservative evolutionary model. Perhaps most relevant at a practical level, the Compara method is intentionally extremely conservative and makes very few inferences compared to the ones an expert GO-PAINT curator would make. This is due to two ad hoc assumptions regarding transfer of function. First, functions are only transferred among vertebrate species; for human, this means only mouse, rat and zebrafish annotations are transferred, and none from the other model organisms. As a result, inferred annotations for human genes are generally lacking for key processes such as DNA replication and repair that have been worked out experimentally in model systems like yeast and E. coli. Second, functions are transferred only when there are no gene duplications following the speciation event that separated the two species. Thus, as the ancestor of teleost fish underwent a whole genome duplication the Compara method makes few functional inferences from zebrafish, and likewise there are few inferences even from mouse and rat for rapidly duplicating gene families such as immune system genes and drug metabolism genes.

Another phylogenetic method has been developed (though specifically only for molecular function prediction, which is not relevant for most applications of GO annotations), called SIFTER (Engelhardt, et al. 2011). We discuss it here, as it was the first method aimed at making probabilistic GO function inferences through use of a phylogenetic tree (the "phylogenomic approach" as coined by Eisen)(Wu and Eisen 2008). SIFTER (Engelhardt, et al. 2011) uses the tree as the structure for "message passing" (the message in this case being a functional annotation) from the leaves to the root, and then back to the leaves, with a particular fidelity determined by a large set of parameters that are determined separately for each protein family. Thus, in essence, the parameters determine how far (in tree distance, i.e. total of all tree branches connecting any two nodes) an annotation from one gene will be shared with other genes. This approach has at least two characteristics that, we contend, make it unsuitable for the general problem of GO annotation inference (note that for genomic data analysis we are primarily interested in annotation of biological pathways and processes). First, SIFTER was developed expressly only for inference of GO molecular function. Indeed, it cannot be applied to process and pathway functions because the number of possible transitions is exponential in the number of functions, so it simply does not scale to solve the problem. Second, the model parameters are transition probabilities among functions, which is appropriate when transitions are frequent but is less appropriate when transitions are rare. Their analogy is to sequence evolution models such as Jukes-Cantor (Erickson 2010) in which multiple transitions typically occur along each branch in the tree, but as work below has shown, function evolution is fundamentally different in that only a few transitions typically occur in the entire tree.

The Gene Ontology Consortium's Phylogenetic Annotation Inference Tool (GO-PAINT) Project was first described in 2011 (Gaudet, et al. 2011), has now reached maturity. This project has, over the past three years, developed a completely unprecedented resource: expert-curated models of gene function evolution for over 4000 distinct gene families. In this project, expert biologist "curators" use a sophisticated software program to view a gene family, experimental GO annotations for different sequences, a multiple sequence alignment, links to information about each gene, and other information. The curators then integrate all this information into a formal model of gain and loss of functions (represented by controlled vocabulary terms from the Gene Ontology) at ancestral nodes in the tree. Thus, for the first time, a large amount of training data is available for designing a scalable computational approach to gene function inference.

As described in the initial publication (Gaudet, et al. 2011), the GO-PAINT Project uses a different formulation for the function inference problem than previous approaches. This formulation treats function as a "character" whose evolution can be reconstructed; the ancestral reconstruction of character gain and loss events can then be used to infer the functions of related genes. Curators of the GO-PAINT project ask: given a set of observed annotations for leaf genes, what is the history (or set of histories) of functional gain and loss events that best explains these observations? The formulation is related to the problem of ancestral character state reconstruction in phylogenetics, for which there is an extensive literature (Cunningham, et al. 1998; Joy, et al. 2016; Pagel 1999). Essentially, human curators attempt to infer a parsimonious model for explaining the distribution of experimentally-determined gene functions, in terms of gain-of-function and loss-of-function events over branches of an evolutionary tree. PAINT curation is performed in a semi-automatic way: the process begins with automatic integration of experimental annotations to homologous genes in a same phylogenetic gene tree, curators then manually create a reconstruction of the evolution of gene function within the family based on the integrated information, the tree structure, sequence features and biological training. This reconstruction explicitly captures inferred functional gain and loss events in specific branches of the tree, which are then propagated to uncharacterized descendent genes.

This human curator-based approach has been crucial for defining the overall process for modeling functional evolution, especially in determining the other data (in addition to the experimental GO annotations, which are very sparse in nearly all gene families) that are required to construct specific models of function evolution across a broad range of families. While PAINT curation requires substantial manual input from trained biocurators, the curators have followed established guidelines to ensure both accuracy and tractability on a large scale.

We present here a probabilistic function inference methods that models the same functional evolution model as pioneered and validated by manual model building for over 4000 gene families in the GO-PAINT project. This model is based on evolutionary reconstruction of ancestral gene functions, which is fundamentally different from any previous work done in the area of computational function inference. Intuitively, as convergent evolution of the same function is rare, a parsimonious model will usually specify that a function was gained prior to the common ancestor of genes for which the function has been experimentally observed, though it may have been subsequently lost in some descendants. Of course, we cannot know for certain if these manual reconstructions of function evolution are correct, but they provide a large set of inferences with which to develop, improve, and validate our inference methods. To some extent, the manual procedures are subjective and subject to variability due to a range of factors such as the completeness of the annotations and differences in curators' expertise. To ensure consistency and reproducibility, detailed guidelines are enforced for PAINT annotations.

As a straightforward way to construct a probabilistic function inference tool, we aimed to develop a model of curators' "subjective" steps in determining which branches to annotate gain-of-function and loss-of-function events. Specifically, we propose a "traceback" model for gain-of-function annotations, followed by a "traceforward" model for loss-of-functions annotations. The "traceback" model is based on the initial (parsimonious) hypothesis that a gene function that has been experimentally demonstrated by multiple extant genes was inherited from their common ancestor. The curator (and algorithm) can then decide between three options: 1. the gain-of-function occurred exactly along the branch leading to the common ancestor of experimentally annotated genes; 2. the gain-of-function occurred prior to the common ancestral branch; 3. there were actually multiple gain-of-function events (convergent evolution) that should be annotated along branches descending from the common ancestor. The second option would follow a "trace-up" route of possible locations for the gain-of-function, which is a consecutive list of ancient ancestors from the MRCA to the root of the tree. This model assumes curators would check these ancient ancestors sequentially, and "stop" at the most likely ancestor. The "traceforward" loss-of-function model assumes curators consider losses during the forward propagation process: for each GO term that has been previously annotated with gain annotations, all descendants of the internal node which has been annotated could potentially inherit the same function; branches with "divergent" nodes, especially those following a gene duplication event, are examined to see if these nodes should NOT inherit the function. If so, a loss-of-function is annotated to specific branches. We fit the "traceback" model and the "traceforward" models separately using the annotations of gain-of-function and loss-of-function events from the manually curated 4000 gene families of the GO-PAINT, and find parameters that best predict the decisions of curators in the "traceback" and "traceforward" routes.

## Materials and Methods

### PAINT annotations and associated info

PAINT curation is the process of annotating evolution events to ancestral genes in phylogenetic genes trees. Currently reference trees in PANTHER database are annotated by PAINT, which include protein-coding genes from all the 12 GO Reference Genomes, and 104 genomes covering all major taxa of species (Mi, et al. 2017). Gene duplication and speciation events are predicted from the phylogenetic trees that are built using a software package GIGA (Thomas 2010), giving biological meaning to internal nodes of the trees and allowing them to be related to the species tree and taxon constraints. The nodes that are predicted as a speciation event stands for an ancestral gene at a specific evolutionary history period, and a duplication node stands for an ancestral duplication event which adds extra copies of ancestral genes to the genome of an ancestral species (Tang, et al. 2018b). The duplicated copies of genes are represented by duplication node's direct descendants which are usually speciation nodes, the time of duplication event is predicted by the oldest period of these descendants. GIGA also estimates branch lengths leading to both leaves and internal nodes, which reflects the number of changes or "substitutions" per site.

The phylogenetic trees are continuously updated with new gene products from more species; phylogenetic gene tree sizes are getting larger, additional new phylogenetic trees are constructed and new trees may also change the tree topology relative to previous versions. There are 13096 phylogenetic gene trees for 130 species in the latest version of PANTHER (PANTHER 11.1, released October 2016), compared with 5422 gene trees for 48 species (PANTHER 7) back in 2009 when PAINT was first launched. Besides, the Gene Ontology terms are updated at a high speed, there were only 35000 GO terms when the PAINT tool was published in 2011 (Gaudet et al., 2011), there are more than 45000 terms in the latest version of GO (The Gene Ontology 2017) (April 2017). In addition, the total number of experimental annotations increases tremendously from only 375000 in 2011 to more than 695000 today (http://amigo.geneontology.org/amigo/search/annotation?q=*:*&fq=evidence_subset_closure_label:%22experimental%20evidence%22&sfq=document_category:%22annotation%22).

The PAINT project has been automatically updating phylogenetic annotations by checking if each GO term has been obsoleted and the validity of each annotation through examination of changes in phylogenetic tree structure, existence of experimental evidences, taxon constraint of GO terms and so on. However correctly inferred phylogenetic gene trees and experimental annotations to current genes are the essential evidence for curators to manually reconstruct evolutionary model of gene functions. Thus, to ensure correct modeling, we collect GO-PAINT annotations along with exact version of the reference PANTHER tree file, the exact set of experimental annotations, which were utilized by curators when they made their decisions.

In this study, we have downloaded the PAINT annotations, the exact sets of supporting materials used by curators to make annotations from

http://viewvc.geneontology.org/viewvc/GO-SVN/trunk/gene-associations/submission/paint/ (date June 2016). These supporting materials include phylogenetic gene tree, experimental annotations and evidence code to extant species genes, the map file that links extant species genes to leaves of the gene tree, and evolutionary inferences of internal nodes of the gene tree.

## Taxon constraints of Gene Ontology terms

Although Gene Ontology terms are designed to be species-agnostic to enable annotations in all species across taxon, some GO terms are limited to certain taxa. For example, "heart" and "heart" related terms should be constrained to animals, which have heart, and should never be allowed to annotate genes of other species like plants, fungi and others. It may seem trivial to PAINT curators who are experiences biologists, but algorithms have no biological knowledge. While building models for this project, we realized that biological knowledge is critical for curators to make assertions about which ancestral gene starts a certain new function. The previous version of taxon constraints of GO terms was built back in 2010, which covers only 599 GO terms (Deegan nee Clark, et al. 2010). We built an updated version of taxon constraints from a combination of manual curation and automatic propagation from constraints of various sources. The taxon constraints are freely available at https://github.com/haimingt/GOTaxonConstraint (Tang, et al. 2018a).

In brief, this new version of GO taxon constraints defines the time(s) of emergence ("gain") and loss of a function in evolutionary history. For example, to express the constraint that a gene can function in the "nucleus" only in eukaryotes, we would express this as "the nucleus emerged on a particular branch of the tree of life, namely the one following LUCA (the last universal common ancestor) and the last common ancestor of all eukaryotes." We express a "never_in_taxon" type of constraint in terms of when a function is lost along a particular branch of the tree of life. When a GO term with taxon constraint is been annotated, the taxon constraint of emergence sets the upper limit of "trace-up" routes for the ancestral nodes which could potentially be annotated with a GO term. While for a GO term without taxon constraint, the endpoint of the trace-up route is the root of the tree.

Taxon constraints are also used to determine if a convergent evolutionary model should be utilized. When the MRCA of a GO term corresponds to an ancestral species that is more ancient than the time of emergence of the GO term, the convergent evolutionary model is more likely: the same function has evolved in several separate sets of genes in evolutionary history instead of having been inherited from one common ancestral gene.

## GO annotations predicted by Ensemble Compara

Ensemble Compara annotations are downloaded from Gene Ontology Annotation (UniProt-GOA) Database (Huntley, et al. 2015) through QuickGO webserver (Binns, et al. 2009) with filter "Ensemble". A total of 6822861 annotations are found as of March 2017. The annotations are then compared with GO-PAINT annotations.

# The "traceback" model for Gain annotation of functions

After extensive examination of the PAINT curator guidelines documentation (http://wiki.geneontology.org/index.php/PAINT_SOP), and in-person discussion with several experienced PAINT curators, we formalized a "traceback" model for gain-of-function annotations of GO terms. The procedure is illustrated in Figure 1. Pseudo-codes for this algorithm is listed in Algorithm 1 in Supplemental Materials.

This algorithm closely follows the standard procedures for PAINT annotations. For each family, we first collect the set of current genes in the reference gene tree that have experimental evidence of this GO term, and find the most recent common ancestor (MRCA) of these genes. This is based on the parsimony assumption: in evolutionary history, new gene functions are difficult to evolve, thus genes with the same function have inherited the function from the same common ancestor. This assumption is most likely to be correct if the taxon constraint for this GO term is older than the corresponding species of MRCA, indicating no conflict in biological knowledge. For example, for GO term "heart morphogenesis", the taxon constraint is "Bilateria"; we have experimental evidence from a mouse gene Tgfb1 and a human gene TGFB2, the most recent common ancestor in PANTHER11 is node PTN000218724 which is predicted to an ancestral duplication event in vertebrates. Thus, there is no conflict with biological knowledge. If the taxon constraint is younger than the species of MRCA, it indicates a conflict. In the previous example, if the species for the MRCA is "Eukaryota", curators know that "heart morphogenesis" only occurs in animals, and couldn't occur in other Eukaryotic species like plants and fungi. Then they will consider a "convergent evolution" model where this function evolves multiple times in history. The alternative reasons for this scenario are addressed in detail in discussion part.

We then find the correct internal node in the reference tree for the species of taxon constraint, and a "trace-up" route is established between the MRCA and this internal node. For the convergent evolution model, we divide the original set of genes with experimental evidences into subsets based on the tree phylogeny and the taxon constraint species. For each subset, a separate MRCA and the internal node for taxon constraint species are found, and the "trace-up" procedure is performed. In this model, the internal node associated with the species of taxon constraint must be deeper in tree structure than MRCA and could be at the root of the reference tree.

We find that in curated models the STOP points in the "traceback" routes appear much more frequently after duplication nodes than speciation nodes. Details are in the results part. This finding is consistent with the hypothesis that new gene functions often evolve after a duplication event which adds an extra copy of the gene to the genome, and thus frees the original copy from negative selection constraints on its evolution (Feuermann, et al. 2016). Thus we model whether STOP decisions (that annotate the GO term to the duplication node's direct child in the traceback route) should be made whenever a duplication node is encountered in the traceback route.

Thus, our model pays special attention to internal nodes inferred with duplication events. During the "trace-up" procedure, when a duplication node is reached, decision is made as to whether the "trace-up" procedure should continue or terminated. If a "CONTINUE" decision is made, the "trace-up" procedure will continue to the next duplication node and another decision will be evaluated. Otherwise, a "STOP" decision ends the "trace-up" procedure, and the GO term under investigation will be annotated to the direct descendent of the duplication node in the trace up route. If the "trace-up" procedure continues to the end of the route, we annotate the GO term to the last internal node of the route. For the convergent evolution model, annotation of the GO term is made for each "trace-up" route.

Notably, the "STOP" decision should be differentiated from "STOP at" decision. The later annotates the GO term to the duplication node instead of the duplication's direct descendant. We'll describe the difference and why we don't use the "STOP at" decision in the discussion part.

## Gathering data on curator decisions for traceback

The decisions of the "trace-up" routes are modeled using a collection of the observed "decisions" through the hypothesized "trace-up" routes for each annotated GO term.

The data gathering procedure is a simulated implementation of the automated annotation procedure, with a slight difference in determination of the "trace-up" route. In the annotation procedure, the trace-up route is from the MRCA to the internal node as indicated by the taxon constraint species of the GO term. In the data mining procedure, the "trace-up" route starts with the most recent common ancestor and ends with internal node annotated by the curator, the duplication nodes in the "trace-up" routes are recorded, the duplication nodes recorded as "PASS" events, and if the parent of the internal node annotated by the curator is a duplication node, then this duplication node is recoded as "STOP after" event. For details, please refer to Algorithm 2 in supplemental material.

The convergent evolution model is considered if MRCA is not a child of the internal node annotated by curator and if subsets of genes with experimental evidences could be determined by the taxon constraint species and the tree phylogeny. A similar trace-up route is repeated for each of the subsets of genes.

A series of parameters are recorded, including the total number of genes in the PANTHER family, the aspect of the Gene Ontology term (molecular function, biological process, or cellular component), the number of direct descendants of the duplication node, the common ancestor species of the duplication node, the branch length of the direct descendant in the trace up route, the ratios of this branch length over sibling's longest and shortest branches, and the among all branches, the ratio of the longest branch over the shortest one, the total number of duplication nodes from MRCA to the root of the tree, the number of passed duplication nodes in the trace-up route from MRCA to this duplication node. Most of the parameters above are suggested by curators as being utilized by them in the manual curation process.

## Regression for the "traceback" model

The binary outcome is the events of "CONTINUE" (0) and "STOP" (1) collected during the trace up procedures using the data mining procedures described above. We chose the simple logistic regression model as a first attempt to check the significance of each parameter. Statistical analysis is performed using Stata Statistical Software: Release 13. College Station, TX: StataCorp LP.

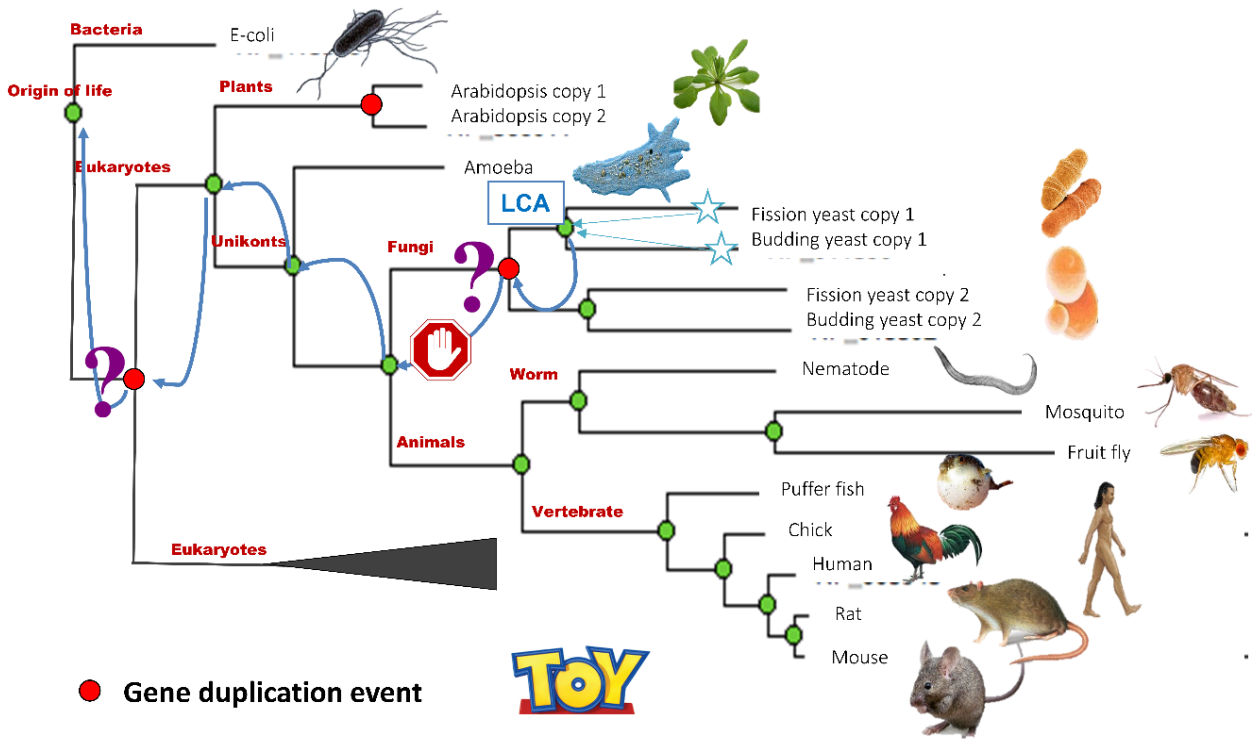

**Figure 1 Illustration of "traceback" model**

*This figure shows an exemplary "trace-up and stop" procedure for annotating a GO term along the phylogenetic gene tree for a protein family. The leaves are orthologous genes of different extant species. These leaves are joined together forming a hierarchical tree structure with speciation events: green dots and duplication nodes: red dots. The blue stars show that the genes that are annotated with a specific GO term. For the first step of "trace-up and stop" model, the most recent common ancestor (MRCA) of the leaf genes that are annotated with this GO term is found. Then a trace-up route is established from this MRCA to the root of the tree as shown by blue arrows in the figure. We then assume a curator to make decisions as to which ancestral node to annotate this GO term through the "trace-up" route. We consider 2 types of "stop"s: (1) continuation of tracing up will violate the taxonomic constrictions of the GO term. This scenario is shown by the "stop" sign along the trace-up route. In this example, the taxon*

*constraint for this GO term is "Fungi", thus the trace-up procedure should stop at or before the internal node for Fungi. (2) A decision is made as to whether to continue tracing or stop whenever a duplication node is come crossed. These decision points are labeled with purple question marks.*

There are a total number of 7731 events collected, with 2663 (36.34%) "STOP" events and 5068 (65.56%) "CONTINUE" events. Summary statistics for the data is in Table 1, detailed descriptions are in the results part. The data points are divided into 10 parts, with 9 parts used for building a prediction model, and the model is applied to the one part left for model validation. A ROC curve is plotted from the predicted p values through application of the logistic regression model to the validation set.

The logistic regression model is used as the "GAIN_decision_model" in Supplemental Algorithm 1.

## "Traceforward" model of loss-of-function events

An ancestral (or extant) gene is annotated with loss annotations to indicate loss of function in evolution. Loss annotations are inherited by descendants like Gain annotations, preventing propagation of gain annotation of the corresponding function to descendants of the internal node annotated with loss function. There are 3 types of gene loss annotations in Gene Ontology annotations: IBD (Inferred from Biological aspect of Descendent) Loss, IKR (Inferred from Key Residues) and IRD (Inferred from Rapid Divergence). IBD is used only if there are "NOT" experimental evidences indicating that certain species cannot have certain function. There are only 6 IBD loss annotations, so we did not initially build models for it. IKR is used to annotate sequences for which absence of specific key residues is observed. e.g. a missing active site residue. Utilization of this code requires curators to check specific sequences in multiple sequence alignments. The difficulties with IBD and IKR loss annotations are discussed in more detail in the discussion section below. IRD or inferred from rapid divergence is used during the process of propagation ancestors' functions to descendants, where curators are not confident to propagate the function to clade with long branch length (indicating a large amount of sequence divergence).

## "Traceforward" model for Loss annotation inference

We built a "traceforward" model assuming annotators consider losses during the down-propagation process: for each GO term that has been annotated to an ancestral node as a gain-of-function, all descendants of the internal node which has been annotated could potentially inherit the same function; when a duplication node is encountered in the propagation, one or more branches are examined to see if the function could have been lost at that point. If so, loss-of-function is annotated to specific branches. The "traceforward" model is illustrated in Figure 2.

Not all duplication nodes in the propagation routes are considered as "potential" loss points. Only those that satisfy 2 requirements are considered: 1. At least one descendant branch has genes with experimental evidence of the GO term under investigation; 2. At least one other descendant branch has no genes with experimental evidence of the specific GO term. This is because only the descendant branch with no experimentally annotated genes could potentially lose the function if it shows much difference with the descendant branch that has genes with experimental evidence.

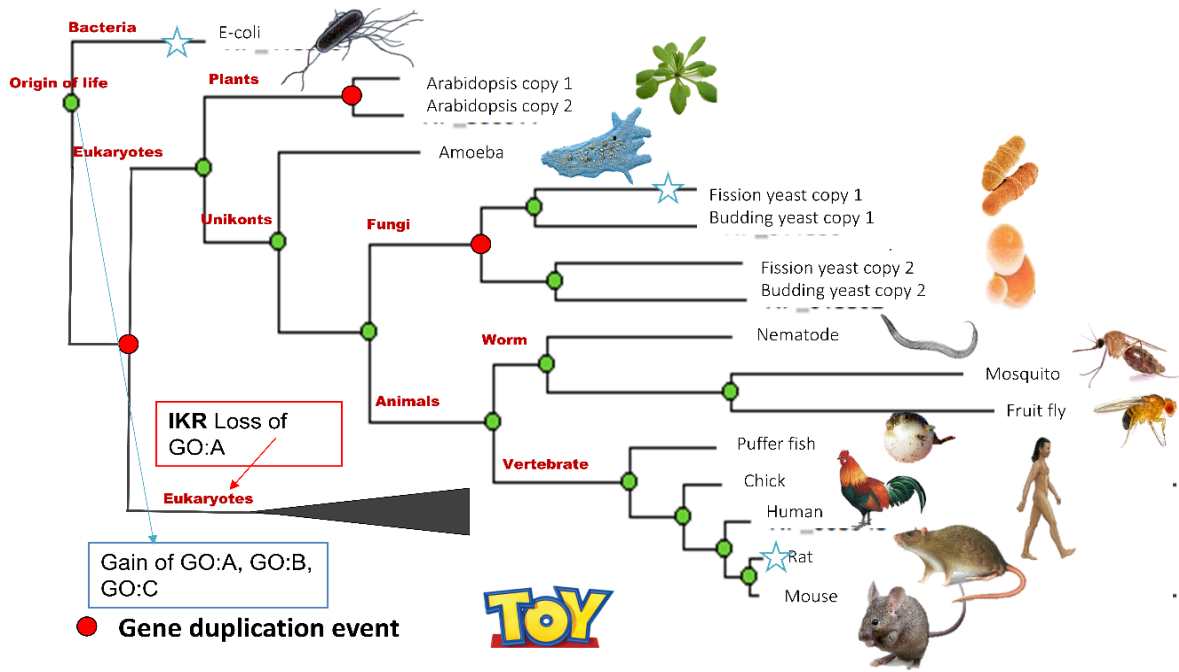

**Figure 2 Illustration of finding potential duplication nodes for IRD loss**

*After applying the "traceback" model for Gain-of-function annotations, the root node of the tree is annotated with several different questions GO:A, GO:B and GO:C as shown by the blue box and arrow. A "traceforward" model for loss-of-function annotations is then applied for each of these GO terms. The propagation process is from the annotated internal node to its decedents until reach the leaves of the tree. In the propagation route, duplication nodes are checked if they are "potential" to be annotated with loss of functions.*

We collect all "potential" duplication nodes for IRD loss annotations during the hypothesized propagation processes of all gained GO terms of all curated families, and recode them as "WITH IRD" events if one or more descendants of the duplication nodes are annotated with IRD, and "WITHOUT IRD" events if no descendants are annotated with IRD.

Several parameters are collected at the same time: the type of the GO term being propagated, the species corresponding to the duplication node, the number of direct descendant branches with experimental evidences, the number of direct descendant branches without experimental evidences, the average branch length of the branches with experimental evidences, the average branch length of the branches without the experimental evidences and the total number of a parameter we named "true negative node". The number of "true negative nodes" adds up 1 if each of the descendant branches has experimental annotations of GO terms other than the GO term being propagated. We use this parameter to check why certain descendant branches have no gene with experimental annotations. A very common situation is that these descendant genes are neither experimentally annotated with the GO term under investigation nor other GO terms in this phylogenetic family, then it is highly likely that these genes are never experimentally studied, thus we do not treat the lack of experimental annotations as potential loss of functions. The other situation is worth more attention: the descendant genes have experimental annotations of GO terms other than the GO term under propagation, but none of these genes are annotated with the GO term under propagation. As these genes are well studied, the lack of experimental annotations is highly likely due to function losses.

## Regression model for loss-of-function events after duplication nodes

A simple logistic regression model is used to fit a prediction model for loss annotations after duplication nodes. The binary outcome is the events of "WITHOUT IRD" (0) and "WITH IRD" (1) collected during the propagation procedures for each previously annotated GO term. The model parameters are as described above. Statistical analysis is performed using Stata Statistical Software: Release 13. College Station, TX: StataCorp LP.

We have found 1171 families with "potential" duplication nodes for IRD loss annotations, only 131 (11.19%) of them have a least one IRD annotation. Thus we fit the prediction model using the data points of the 131 families. There are 2412 events collected, with 267(11.07%) "WITH IRD" events and 2138(88.64%) "WITHOUT IRD" events. Summary statistics for the data is in Table 2, detailed descriptions are in the results part. The data points are divided into 10 parts, with 9 parts used for building a prediction model, and the model is applied to the one part left for model validation. A ROC curve is plotted from the predicted p values of the data points in the part left.

The final prediction model is used as the "IRD_decision_model" in Algorithm 3.

## The complete process of the automation tool

The complete process of PAINT automation is a combination of the "traceback" model for gain-of-function annotations and "traceforward" model for loss-of-function annotations, followed by propagation process of gain-of-function annotations, during which loss-of-function annotations stop the propagation process. Details for the propagation process is listed in Algorithm 3 in supplemental material.

## Select the GO terms to be annotated

Not all GO terms with experimental annotations are used for PAINT annotations, curators aim to annotate the most specific and appropriate term. Out of the 17385 different GO terms that have experimental annotations, only 6950 (39.98%) GO terms are used for annotations in the gaf files. Thus, it is a necessity to prioritize the GO terms to be annotated from all GO terms with experimental evidences. However, this is a very complex problem as different curators often have different opinions about the most "appropriate" term.

To take a first step toward addressing this problem, we constructed a simple logistic regression model for the possibly of GO terms being selected for annotation by curators. The binary outcome is assigned with 1 if a GO term with experimental evidence is always annotated, and 0 if it is never annotated. The GO terms that are partially selected for annotation are not included in this model: the GO term are selected for annotation in certain PANTHER families but are not selected in other families. We then collected several features for each GO term: the GO type, GO term specificity which is represented by the total number of steps from this GO term to root in GO's acyclic graph and the second last parent of the GO term (the direct child of one of the root terms: biological process, molecular function and cellular component).

# Results and discussions

## PAINT statistics

In table 1, we summary some statistics for annotated PAINT families. There are 3930 PAINT families that have been annotated, 30.01% of the 13096 total families in the latest version of PANTHER database (Version 11.1). 137 families have some missing data and are excluded from further analysis.

In the 3793 annotated PAINT families, there are 462709 extant genes, 75042 duplication nodes and 256191 speciation nodes. Duplication node represents a duplication event in extant species or in a common ancestor in evolutionary history. Speciation node represents a reconstructed ancestral gene of a common ancestor. The 173974 experimental evidences on 17385 different GO terms and 30661 different gene products are mainly from 12 reference genomes, 19760 annotations on 6950 different GO terms are curated on 74 common ancestors, and are then propagated to the 104 current genomes in the next step.

The loss annotations are much fewer than gain of function annotations (651 compared with 19760). Most of these loss annotations have IRD or IKR codes, with only 6 annotations with IBD code which indicates the annotations were made based on the "loss of function" experimental annotations.

Table 1 Summary statistics of GO-PAINT (until June 2016)

| | | Total numbers | |
|---|---|---|---|
| Panther family info in PAINT | PANTHER families with PAINT annotations | 3930 | |
| | skipped families | 137 | |
| Nodes info of Panther families | extant genes in PANTHER families (leaves) | 462709 | |
| | Duplication nodes in PANTHER families (including root) | 75042 | |
| | Speciation nodes in PANTHER families (including root) | 256191 | |
| | Horizontal transfers | 1354 | |
| | species in PANTHER families | 213 | |
| | Species in experimental evidences | 79 | |
| | | Total number of different GO terms or genes | Total number of associated GO annotations |
| Experimental evidences | GO terms with experimental annotations GAIN | 17385 | 173974 |
| | Genes with experimental annotations GAIN | 30661 | 173974 |
| | GO terms with NOT experimental annotations | 714 | 1079 |
| | Genes with NOT experimental annotations | 1079 | 1079 |
| PAINT annotations | GO terms used for "Gain" PAINT annotations | 6950 | 19760 |
| | GO terms used for "Loss" PAINT annotations | 514 | 651 |
| | Genes used for Gain annotations | 20859 | 46119 |
| | Genes used for Loss annotations | 2261 | 3161 |

## Relationships between PAINT annotations and node types

We look at PAINT annotations for 6 distinct types of nodes:

"node is duplication, parent is duplication", "This is duplication, parent is speciation", "This is speciation, parent is duplication", "This is speciation, parent is speciation", "This is root of a tree, and is speciation" and "This is root of a tree, and is duplication". Specially, "This" here stands for the internal code under investigation; "parent" is this internal node's direct parent. If "This" is the root of a reference tree, then it has no parent. The 6 scenarios are the total possible combinations of This and parent's node types: speciation node or duplication node.

Then we count the number of nodes, which are annotated with gain-of-function, or loss-of-function for each of the 6 scenarios, together with the total number of annotated GO terms and the number of annotations. Detailed results are listed in Table 2. Two way tables are constructed for node types and the number of nodes with or without annotations, separately for only gain-of-function, only loss-of-function and both gain-of-function and loss-of-function annotations; Pearson Chi-squared is used for calculating the association. Detailed results are listed in SupplementalTable1.1, 1.2 and 1.3.

From the results, distinct types of nodes have statistically different probabilities to be annotated with gain of function or loss of function (both $p<0.001$). We see speciation nodes whose parents are duplication nodes and both types of root nodes are more likely to be annotated with gain of functions compared with other types of nodes. Speciation nodes whose parents are duplication nodes are also more likely to be annotated with loss of functions. These observations are supportive of our decision to "STOP" duplication nodes in the "traceback" routes for gain function annotations, and the decision to annotate IRD loss to descendant of "potential" duplication nodes in the propagation routes.

Gene duplication has been widely hypothesized to be an opportunity to evolve new functions and/or lose ancestral functions. To test this hypothesis, we check the scenario when both gain and loss of function are annotated to one or more descendant of "This" internal node. If "This" node is a duplication node (its parent may be a duplication, a speciation node, or it is the root of the tree), then its descendants are statistically more likely to be annotated with both gain and loss of functions (87.21% of 219 cases, $p <0.001$).

## The "traceback" annotation model

### Summary statistics

Summary statistics of the "traceback" model for gain annotations are listed in Supplemental Table 2. There is a total of 7731 events collected in the "traceback" routes, 2663 (34.46%) out of them are "STOP" events, and the rest 5068 are "CONTINUE" events. The "STOP" events involve 2013 different GO terms and 701 different PANTHER families, while the "CONTINUE" events involve 1990 different GO terms and 1432 different PANTHER families. Each different common ancestor species for the duplication nodes is statistically significant for different types of events (p<0.001). There are 46 distinct species for the "STOP" and 57 "CONTINUE" events. 4 types of Gene Ontology terms are collected: biological process, cellular component, molecular function, and metabolic molecular function if a molecular function term represents a metabolic process. Types of GO terms are statistically significantly associated with event types (p=0.006), with more biological process terms in "STOP" events than "CONTINUE" events, and more other types of GO terms in "CONTINUE" events. Duplication nodes in the trace-up route that have more than 2 direct children (32.32%) are more likely to be "STOP" (p<0.001): 41.46% of the "STOP" duplication nodes have more than 2 direct children compared with 27.53% for that of "CONTINUE" duplication nodes. The average branch length of the direct descendant of the duplication nodes in the trace-up routes is 0.56±0.49 (mean±SD), the branch length of the direct descent of "STOP" duplication nodes are statistically longer than that of "CONTINUE" duplication nodes (p<0.001). The average number of genes in PANTHER families is 386.71±534.86 (mean±SD), with this

**Table 2 Association of nodes type and PAINT annotations**

| | Nodes types | # nodes | # nodes whose children are annotated with both Gain and Loss | # nodes with Gain or loss annotations | Annotation type | # annotated GO terms | Percentage of GO terms among all annotataed GO terms | # Annotations | Percentage of GO terms among all annotataed GO terms |
|---|---|---|---|---|---|---|---|---|---|
| PAINT annotations by nodes type | Mom is duplication, this is duplication | 2727 | 3 | 16 | Gain | 43 | 0.62% | 43 | 0.02% |
| | | | | 4 | Loss | 6 | 1.17% | 6 | 0.92% |
| | Mom is duplication, this is speciation | 39464 | 16 | 2984 | Gain | 3481 | 50.09% | 6645 | 3.82% |
| | | | | 363 | Loss | 460 | 89.49% | 715 | ~ |
| | Mom is speciation, this is duplication | 71278 | 173 | 298 | Gain | 484 | 6.96% | 648 | 0.37% |
| | | | | 22 | Loss | 47 | 9.14% | 62 | 9.52% |
| | Mom is speciation, this is speciation | 212677 | 11 | 1427 | Gain | 1518 | 21.84% | 2566 | 1.47% |
| | | | | 106 | Loss | 74 | 14.40% | 135 | 20.74% |
| | Annotation to root, root is duplication | 705 | 15 | 550 | Gain | 1009 | 14.52% | 1805 | 1.04% |
| | | | | 0 | Loss | 0 | 0.00% | 0 | 0.00% |
| | Annotation to root, root is speciation | 3127 | 1 | 2669 | Gain | 3518 | 50.62% | 9287 | 5.34% |
| | | | | 1 | Loss | 1 | 0.19% | 1 | 0.15% |

number statically larger in families with "STOP" duplication nodes (p<0.001). 3 types of branch length ratios among duplication node's direct descendants are summarized: ratio of the branch length of the descendant in the trace-up route over sibling's longest branch length, ratio of the branch length of the descendant in the trace-up route over sibling's shortest branch length, and ratio of the longest branch length over the shortest branch length among all descendants. These

3 ratios are statistically significantly associated with decision events (p< 0.001, p<0.001 and p=0.005). Each of 3 ratios is smaller for the "STOP" duplication nodes.

**Prediction model building for "traceback" model and model validation**

Table 3 lists details of the prediction model based on the randomly choose 9 parts of the whole data set, the predictors include branch length of the direct descendant of the duplication node in the trace-up route, number of children of the duplication node, number of genes in the PANTHER family, ratio of the branch length of the descendant in the trace-up route over sibling's longest branch length, ratio of the longest branch length over the shortest branch among all descendants, number of passed duplication nodes along the trace-up route, the ratio of the passed duplication nodes over the total number of duplication nodes from the most recent common ancestor to root, and the common ancestor species for the duplication node. The prediction model is applied to the validation data set, an ROC curve is plotted using the predicted p values. The area under curve for the ROC curve is 0.753, which indicates an acceptable prediction model.

**Table 3 Logistic regression model results for "traceback" model**

| Logistic regression for PASS through or STOP at a duplication node | | | | Number of obs | 6797 | |
|---|---|---|---|---|---|---|
| | | | | LR chi2(50) | 1137.36 | |
| | | | | Prob > chi2 | 0 | |
| STOP =1 | Odds Ratio | Std. Err. | z | P>z | [95% Conf.Interval] | |
| Branch length | 2.02836 | 0.135552 | 10.58 | 0 | 1.779347 | 2.31222 |
| Number of children | 1.365071 | 0.089398 | 4.75 | 0 | 1.200633 | 1.552031 |
| branch ratio 1 | 0.939783 | 0.012604 | -4.63 | 0 | 0.915401 | 0.9648144 |
| branch ratio 2 | 1.044362 | 0.011832 | 3.83 | 0 | 1.021428 | 1.067811 |
| Number of genes | 1.000433 | 0.000062 | 6.99 | 0 | 1.000312 | 1.000555 |
| passed Duplications | 0.4123614 | 0.023771 | -15.37 | 0 | 0.368307 | 0.4616859 |
| passratio | 0.3365609 | 0.038518 | -9.52 | 0 | 0.268935 | 0.4211914 |
| Speciation | Not shown | Not shown | Not shown | <0.0001 | Not shown | Not shown |

## The traceforward loss-of-function model

### Summary statistics

Summary statistics of the "traceback" model for gain annotations are listed in Supplemental table 3. There is a total of 2412 duplication nodes collected in the "propagation" routes, 267 (11.07%) out of them are "WITH IRD" loss annotations in direct descendants, and the rest 2138 are "Without IRD" loss annotations. The "With IRD" events involve 193 different GO terms and 131 different PANTHER families, while the "Without IRD" events involve 518 different GO terms and 112 different PANTHER families. Each different common ancestor species for the duplication nodes is statistically significant for different types of events ($p<0.001$). There are 48 distinct species for the "With IRD events" and 30 "Without IRD" events. 4 types of Gene Ontology terms are collected: biological process, cellular component, molecular function, and metabolic molecular function if a molecular function term represents a metabolic process. Types of GO terms are statistically significantly associated with event types ($p=0.0011$), with more molecular function and metabolic molecular function terms in "With IRD" events than "Without IRD" events, and more other types of GO terms in "Without IRD" events. The number of descendants with experimental annotations of other GO terms is differently distributed between "With IRD" and "Without IRD" ($p<0.001$). But opposite to what we have thought, most of "With IRD" events associated duplication nodes have no descendants with experimental annotations of other GO terms (82.02%), compared with 40.41% for that of "Without IRD" events. The number of direct descendants with Gain annotations is differently distributed between "With IRD" and "Without IRD" ($p=0.004$), "With IRD" events associated duplication nodes are more likely to have only 1 such descendant branch. The number of direct descendants without Gain annotations is also differently distributed between "With IRD" and "Without IRD" ($p=0.001$), "With IRD" events associated duplication nodes are more likely to have 3 or more such descendant branches without Gain annotations. The average branch length of the direct descendants with Gain experimental annotations of the potential duplication nodes in the propagation routes is 0.36±0.43 (mean±SD), the branch length of the direct descendants of "With IRD" duplication nodes are statistically longer than that of "Without IRD" duplication nodes ($p<0.0012$). The average branch length of the direct descendants without Gain experimental annotations of the potential duplication nodes in the propagation routes is 0.51±0.52 (mean±SD), the branch length of the direct descendants of "With IRD" duplication nodes are statistically longer than that of "Without IRD" duplication nodes ($p<0.001$). The average ratio of the branch length with gain experimental annotations over the branch length without gain experimental annotations is 2029±11.22 (mean±SD), this ratio of "With IRD" duplication nodes are statistically longer than that of "Without IRD" duplication nodes ($p<0.001$).

## Regression model building for "traceforward loss" model and model validation

Table 4 lists details of the prediction model based on the randomly choose 9 parts of the whole data set, the predictors include the common ancestor species for the duplication code: Dup_species, the number of direct descendants with Gain experimental annotations of the "potential" duplication nodes in the propagation routes: branch_Gain, the average branch length of these branches with Gain annotations: Gain_length, the number of branches without Gain experimental annotations: branch_no_Gain, the average branch length of the branches without Gain annotations: No_Gain_length, and the number of branches which are annotated with GO terms other than the GO term under investigation: trueN. The prediction model is applied to the validation data set, an ROC curve is plotted using the predicted p values. The area under curve for the ROC curve is 0.903, which indicates a good prediction model.

**Table 4 Logistic regression model results for "traceforward" model**

| | | | | Number of obs | | 2139 |
|---|---|---|---|---|---|---|
| | | | | LR chi2(50) | | 542.71 |
| Logistic regression for if IRD should be annotated to one or more descendants of a "potential" duplication node in the propagation route | | | | Prob > chi2 | | 0 |
| | | | | Pseudo R2 | | 0.3314 |
| "With IRD" =1 | Odds Ratio | Std. Err. | z | P>z | [95% Conf.Interval] | |
| Dup_species | Not shown | Not shown | Not shown | <0.0001 | Not shown | Not shown |
| branch_Gain | | | | | | |
| 2 | 0.5035908 | 0.147452 | -2.34 | 0.019 | 0.283691 | 0.893944 |
| 3 | 0.5499776 | 0.214694 | -1.53 | 0.126 | 0.255896 | 1.182025 |
| branch_no_Gain | | | | | | |
| 2 | 2.009275 | 0.52559 | 2.67 | 0.008 | 1.203318 | 3.355046 |
| 3 | 44.73397 | 15.86001 | 10.72 | 0 | 22.3281 | 89.62373 |
| trueN | | | | | | |
| 1 | 0.0447291 | 0.011691 | -11.89 | 0 | 0.026799 | 0.074656 |
| 2 | 0.0082972 | 0.004229 | -9.4 | 0 | 0.003055 | 0.022532 |
| 3 | 0.0011847 | 0.000813 | -9.82 | 0 | 0.000309 | 0.004547 |
| Gain_length | 0.5387939 | 0.115043 | -2.9 | 0.004 | 0.354547 | 0.818788 |
| No_Gain_length | 2.547261 | 0.427175 | 5.58 | 0 | 1.833695 | 3.538504 |
| _cons | 0.0027168 | 0.002825 | -5.68 | 0 | 0.000354 | 0.020846 |

## Inference validation

We validated our models by comparing the predicted annotations with curators results as well as the Ensemble Compara annotations. Details in Table 5. In PAINT, curators' annotations to internal nodes are propagated to 1755662 annotations made to leaves of the genes. 68.7% of these annotations are predicted by our automation tool while only 21.8% are predicted by Ensemble Compara. Out of the 548959 annotations that are made by curators but not predicted by our automation tool, 21.6% are from "IRD potential loss" annotations made by the automation tool. There are 819389 annotations that are made by our automation tool but not predicted by curations, 1.45% are from "IRD potential loss" annotations made by the curators. The results show that the annotations that are predicted by the automation tool but not predicted by curators are largely resulted from decision points chosen by the automation tool that are closer to the root nodes than the decision points chosen by curators. This automation tool predicts 2026092 annotations, compared with 1755662 predicted annotations by the curators, and only 627657 predictions made by the Ensemble Compara. Ensemble Compara uses an extremely conservative algorithm in that only one to one and apparent one to one orthologies are used, and only manually annotated GO terms with an experimental evidence type of either IDA, IEP, IGI, IMP or IPI are projected. However, the conservativeness does not lead to more consistent results compared to annotations made by curators: 61.1% of the predictions made by Ensemble Compara are consistent with annotations made by curators in PAINT, and a similar percentage of the predictions by the automation tool are consistent while automation tool predicts 2.22 folds more annotations.

**Table 5 Comparisons of PAINT annotations made by curators with inferences of this automation tool and Ensemble predictions**

| Number of Annotation (gene GO term pair) | | Curators' decisions | | |
|---|---|---|---|---|
| | | With annotation | Without annotation | IRD |
| Predictions of this automation tool | With annotation | 1206703 | 807503 | 11886 |
| | Without annotation | 430177 | NA | NA |
| | IRD | 118782 | NA | NA |
| Ensembl Compara predictions | With annotation | 383761 | 243896 | |
| | Without annotation | 1371901 | NA | |

However, determination of "correct" GO annotations is a hard to solve problem. The annotations made by human curators are not necessarily more accurate than the predictions made by automatic pipelines. As suggested by a recent publication(Feuermann, et al. 2016), a potential way to show the validity of the PAINT annotations made by curators is to examine the percentage of predicted annotations by curators that are proved by experimental evidences after the time-point when curators' make annotations. This comparison will be described in more detail by a separate publication.

# Existing problems and future considerations

## Inconsistencies in different curators' annotations

Although curators follow established guidelines, their decisions of annotating the GO terms to ancestral nodes could be subjective and thus lead to inconsistency between different curators. From Supplemental table 1.3, we see most annotations are curated to speciation nodes after a duplication node. However, there are still 298 cases where annotations are curated to a duplication node instead of one of its descendant. Also, there are 1427 cases where annotations are made to a speciation node whose mom is also a speciation node.

In our "traceback" model for gain-of-function annotations, the traceback routes should stop at the root of the tree or a speciation node after a duplication node or endpoint node of the "traceback" route, which is determined by taxon constraint of the GO term. Thus, in this model, a GO term is annotated to a duplication node (like the 298 cases above) or a speciation node's child (like the 1427 cases above) only if the node annotated is the endpoint of the traceback route determined by the taxon constraint of the GO term being annotated. However manual examination of such cases shows that most of these nodes are not endpoints of the trace-up routes by taxon constraint. Thus, we suspect such cases are from inconsistencies of different curators.

## Curators' annotations that do not follow the procedure

We also observe 1360 cases where biocurators' decision points are children of the most recent common ancestor of experimental annotations. We have found several potential explanations for these cases. First experimental annotations are examined by biocurators, and some of them are filtered if they are believed to be invalid. This usually happens to high-throughput experiments, which generate many experimental annotations with one publication. To solve this problem, we have filtered the publications, which contribute more than 40 pieces of experimental evidences. After the filtering of some experimental annotations, if the species associated with the MRCA is older than the taxon constraints species of the GO term being annotated, then we consider a convergent evolutionary model. In that way we reject the original hypothesis that the function has emerged once in evolutionary history and been inherited by its descendants. Instead we believe the function has emerged during several different time periods separately in evolutionary history. We first find internal nodes that are associated the taxon constraint species, and then find the genes with experimental evidences that are children of each of such internal nodes, and run the "traceback" model using the MRCA of these genes as a start point and the internal node as the upper limit point.

## Incompleteness of "IRD/IKR" loss annotations

IRD is the abbreviation for "inferred from rapid divergence", it is usually decided by curators through comparison of the sequences and without support from NOT experimental evidences. IKR is the abbreviation for "inferred from key residues", it is usually decided by curators through manually checking the key residues of sequences in different clades. Most curators are hesitant to make IRD or IKR annotations. Only 11% of the PANTHER families with "potential" duplication nodes for IRD loss in the propagation routes are annotated with at least one IRD or IKR loss annotations. We only used these PANTHER families to build the "IRD/IKR loss" model for Loss annotations. We then apply the model to PANTHER families without IRD/IKR annotations. A lot of "IRD" losses are predicted for these families. We are providing these predicted Loss annotations to curators to review for the validity.

Although uncommon, "IRD" sometimes are not annotated to direct descendants of a duplication node, instead they are annotated to child nodes a speciation node. These annotations usually require substantial examination of the sequence variations and key residual information.

## The difficulty of selecting GO terms to annotate

Selection of GO terms to annotate is a highly subjective and difficult, mainly because of the complexity of GO structures and annotations to various GO terms that are related with each other. Here we present a case study of PANTHER family to illustrate this problem. The exteant genes in PANTHER family PTHR10025 are annotated with a series of related GO terms: GO:0009113(purine nucleobase biosynthetic process), GO:0009396(folic acid-containing compound biosynthetic process), GO:0004487(methylenetetrahydrofolate dehydrogenase (NAD+) activity), GO:0006730(one-carbon metabolic process), GO:0004329(formate-tetrahydrofolate ligase activity), GO:0004477(methenyltetrahydrofolate cyclohydrolase activity), GO:0009257(10-formyltetrahydrofolate biosynthetic process), GO:0046656(folic acid biosynthetic process), GO:0004488(methylenetetrahydrofolate dehydrogenase (NADP+) activity), GO:0006139(nucleobase-containing compound metabolic process), GO:0035999(tetrahydrofolate interconversion), GO:0048702(embryonic neurocranium morphogenesis), GO:0001843(neural tube closure), GO:0006555(methionine metabolic process), GO:0009069(serine family amino acid metabolic process). Biocurators tend to annotate molecular functions, which are usually building blocks of the biological processes that may involve multiple molecular functions of multiple systems. And they tend to annotate those GO terms that are essential and "core" to the gene family. Only 3 of the GO terms above are selected by the curator. GO:0004488(methylenetetrahydrofolate dehydrogenase (NADP+) activity) GO:0004477(methenyltetrahydrofolate cyclohydrolase activity) and GO:0006730 (one-carbon metabolic process. Even when the curator has decided to annotate the core molecular functions, it is hard to choose between these close functions. It is not wrong nor inappropriate to have different selections. It is just that people have their own decision as to which GO terms best reflect the core of the genes in protein families.

Selection of GO terms to annotate is a hard problem to solve. We tentatively solve this problem by differentiating GO terms that are always annotated when there are experimental evidences and GO terms which are never used for annotation. We then group the GO terms to their high-level parent terms in the GO hierarchical structure. We find certain categories are used for annotations much more often than other categories. Results are included in Supplemental Table 4. Examples of often used categories include "catalytic activity", "metabolic process" and "macromolecular complex" and the rarely used categories include "biological regulation", "response to stimulus", "developmental process".

When apply our automation model to new gene families, we annotate all GO terms that are always or sometimes annotated in existing PANTHER families, and don't annotate the GO terms that are never annotated in existing PANTHER families. For GO terms that have never occurred, we apply the regression model built using high-level parent terms as a predictor, and will not annotate GO terms that belong to rarely used categories. In this way, we ensure automatic annotations of as many as potential GO terms as possible. Predictions for these GO terms will be provided to biocurators to decide which GO terms they want to annotate and propagate.

Future considerations could dig deeper into the relationships among GO terms, find rules to determine the "core" GO terms and the GO terms which could be neglected at the presence of other GO terms.

## Imperfections with taxon constraints

Our automation tool relies heavily on taxon constraints for determination of stop points of the trace-up routes and determination if convergent evolutionary model should be used. Thus, the accuracy of the taxon constraints is significant for the robustness of this model. A tremendous amount of work has done to check the correctness of the taxon constraints and improve taxon constraints at the presence of conflicts with PAINT annotations. We have applied the latest version of constraints for annotations using PAINT project of the GOC. If curators try to annotate a gene from a species outside of the "Gain_at" taxon, the system will warn the curators of potential conflicts. They will give us feedback if taxon constraints are incorrect.

## Biases in annotations

Perhaps the most challenging aspect of the gene function inference is not simply the sparseness of the GO annotations, but rather the bias in the annotations. Nearly all the GO annotations are statements about the functions that a gene has; very few GO annotations are statements about the functions that a gene does NOT have. Currently, there are nearly 600,000 experimental annotations in the GO database, of which less than 5000 (less than 1%) were negative annotations. This is not surprising, as in practice it is much easier, and more publishable, to prove something's existence than its nonexistence. This bias is of paramount importance for any functional inference method, manual or computational. To understand why, consider the following thought-experiment. If one only has evidence of functions that a gene has, but not of the functions that the gene does NOT have, then specifying a parsimonious model of function evolution is trivial: one can simply assert that all the functions evolved prior to the common ancestor of the entire family (the root of the tree), and no further function evolution occurred. Using such a model, one would thus predict that all members of the family perform any function that has been observed for any one member. This approach would be disastrous, as it would generally result in a large number of false positive inferences. Thus, it is essential for ancestral reconstruction to augment these very sparse negative annotations with additional information that allows inference of loss or absence of function.

## Potential additional information for the inference methods

For the GO-PAINT Project, human curators have found several types of information to be sources of useful evidence for such inferences. This information falls into four classes: 1) information about evolutionary events such as gene duplication and speciation, 2) annotated protein sequence features, 3) knowledge of biological system evolution and 4) complementarity of accumulated experimental annotations. For the manual GO-PAINT Project, currently all of this information comes from manual inspection.

The model we are presenting here well incorporates the first three classes of additional information. As shown in the results part, information about evolutionary events plays important roles in curators' annotations. Significant differences in curators' annotations are found for speciation v. duplication events ($p < 0.0001$ for both gain-of-function and loss-of-function annotations). Gain-of-function annotations are more likely to be curated to be after duplication nodes, and to root nodes, and less likely than expected to be after a speciation event. Losses of function are also more likely to be curated following gene duplication. These findings form the basis of our "traceback" gain-of-function model and "traceforward" loss-of-function model. We consider decisions of "stop" and "potential loss" to be only after a duplication node in the trace-up routes for "gain-of-function" annotations and propagation routes for "loss-of-function" annotations.

Similarly, we have found that functional gains and losses occur more commonly in some branches of the species tree than at others. The branch information is critical predictor used for both the gain-of-function model and loss-of-function model. Thus, we've identifiable periods and lineages that were "hotspots" for biological innovation during evolution. For instance, the eukaryotic common ancestor is over 10-fold enriched in functional gain events than an average speciation node; this is even more pronounced for cellular component annotations (>15-fold) with the evolution of eukaryotic cellular structure and organization. The vertebrate common ancestor is also about 6-fold enriched in gain events than average, due largely to the evolution of vertebrate-specific biological processes.

One of the major pieces of evidence used by curators is prior knowledge of biological system evolution. Some of this information is stored in the GO database in the form of "taxonomic constraints". A simple example is that only eukaryotic organisms have a nucleus, so one can never infer that, in the common ancestor of all cellular organisms, a particular gene functioned in the nucleus. Another example would be the constraint limiting the Wnt signaling pathway to animals. In effect, such constraints provide implicit negative annotations for all genes from organisms outside the specified taxa. We have built a new version of taxon constraints for most GO terms, which we would resemble expert curator knowledge on which correct decision of the annotations relies.

Databases such as Swiss-Prot (Boutet, et al. 2016) and the Conserved Domain Database (CDD) (Marchler-Bauer, et al. 2015) have recorded extensive information about protein functional sites. GO-PAINT curators use information about evolutionary changes in functional sites to infer branches in the tree where function may have changed. For example, currently a curator reads a Swiss-Prot entry, finds the active site in the family multiple sequence alignment, and finds genes in the tree in which the active site residues have changed during evolution. Enzyme active site residues and protein domains (large functional modules of the proteins that can be gained and lost as a unit during evolution) are statistically the most highly used, but there are a number of annotated features that curators have found useful for this task. Currently this model hasn't incorporated these features. In the future, we will compute the ancestral states and evolutionary changes in these features along the tree, and identify the branches of the tree along which changes have occurred. We can then allow model parameters to depend on these changes as well. This task will be greatly facilitated by the fact that ancestral protein sequences have already been computed for all tree nodes using the PAML software (Yang 1997).

### Integrate the inferred annotations into the Gene Ontology workflow

As noted earlier, the manner in which our annotations will be integrated into the GO-PAINT project workflow will depend on how well our methods are able to recapitulate the manually curated annotations that have already been deposited in the GO database. Our computational methods will be integrated into the Phylogenetic Annotation Inference Tool as a starting point to assist curators make annotations and to assess the improvement in efficiency of curation.

## Conclusion

In this chapter, we present a first pass of the automation tool for gene function inference with a "traceback" model for gain-of-function annotations and a "traceforward" model for loss-of-function annotations which simulate the workflow of biocurators' manual process of evolutionary function model building and get trained using the manually curated and validated 4000 phylogenetic families of GO-PAINT project. The models also combine phylogenetic rules, phylogenetic tree features including lineages and branch length and taxonomic constraints.

The backbone of the method strictly follows the annotation guidelines of PAINT. 2 models are built separately for the annotation of gain function: the "traceback" model which introduces the concept of "traceback" routes from MRCA to the internal node decided by the taxon constraint species of the GO term under investigation and the "traceforward" model for annotation of loss function: after Gain annotations are made, duplication nodes in the propagation routes are examined.

The validity of the model is established through comparison with annotated PAINT families, it has already been integrated to the PAINT tool, with predicted annotations provided to curators. Curators can decide if they agree with the annotations predicted by the automation tool, or they will reject. The automation pipeline will greatly reduce the time and energy spent by PAINT curators, and greatly accelerate the PAINT project.

This automation tool tries to simulate curators' sophisticated scientific decisions. For a first pass, the decision model and implementation of the tool is quite simple. Future developments could incorporate extra source of information like annotated features of sequences. This tool has already been built into the GO-PAINT tool to assist curators make annotations; further modifications could potentially be based on their feedbacks.